\UseRawInputEncoding
\documentclass[aps,amssymb,amsmath,pra,twocolumn,showpacs,superscriptaddress]{revtex4-1}
\pdfoutput=1
\usepackage{graphicx}
\usepackage{dcolumn}
\usepackage{bm}
\usepackage{subfigure}
\usepackage{color}

\usepackage{lineno}
\usepackage[matrix,frame,arrow]{xy}
\usepackage{epsf}
\usepackage{placeins}
\usepackage{babel}
\usepackage{times}
\usepackage{latexsym}
\usepackage{fancyhdr}
\usepackage{float}
\usepackage{afterpage}
\usepackage{enumitem}
\usepackage{eso-pic, graphicx}
\usepackage{natbib}

\definecolor{Ablue}{rgb}{0.96,0.24,0.00}

\definecolor{orange}{rgb}{0.96,0.24,0.00}

\definecolor{darkred}{rgb}{0.55, 0.0, 0.0}

\usepackage[colorlinks=true , citecolor=blue,urlcolor=blue]{hyperref}

\newcommand{\beq}{\begin{equation}}
\newcommand{\eeq}{\end{equation}}
                  
\newcommand{\benum}{\begin{enumerate}}
\newcommand{\eenum}{\end{enumerate}}
                    
\newcommand{\bit}{\begin{itemize}}
\newcommand{\eit}{\end{itemize}}

\newcommand{\bea}{\begin{eqnarray}}
\newcommand{\eea}{\end{eqnarray}}

\newcommand{\bev}{\begin{verbatim}}
\newcommand{\eev}{\end{verbatim}}

\newcommand{\zfl}[1]{\protect\label{fig:#1}}
\newcommand{\zfr}[1]{Fig. \ref{fig:#1}}

\newcommand{\ba}{\left\{ \begin{array}{lr}}
\newcommand{\ea}{\end{array}\right.}

\newcommand{\blist}[1]{
 \begin{list}{#1}
 \begin{align}
	 arrow
 \end{align}
 $\checkmark\star
  { \setlength{\itemsep}{3pt}
     \setlength{\parsep}{2pt}
     \setlength{\topsep}{3pt}
     \setlength{\partopsep}{0pt}
     \setlength{\leftmargin}{1em}
     \setlength{\labelwidth}{1em}
     \setlength{\labelsep}{0.5em} } }
\newcommand{\elist}{
  \end{list}  }

\DeclareMathSymbol{\vartheta}{\mathalpha}{letters}{"12}
\DeclareMathSymbol{\theta}{\mathalpha}{letters}{"23}
\DeclareMathSymbol{\phi}{\mathalpha}{letters}{"27}
\DeclareMathSymbol{\varphi}{\mathalpha}{letters}{"1E}

\newcommand{\bef}
{
\begin{figure}[htbp]
\centering
}

\newcommand{\eef}{\end{figure}}

\newcommand{\affA}{Department of Chemistry, Chair of Physical Chemistry, Technical University of Munich,  Lichtenbergstra{\ss}e 4, 85748 Garching bei M{\"u}nchen, Germany}
\newcommand{\affB}{Munich Center for Quantum Science and Technology (MCQST), Schellingstr. 4, D-80799 M{\"u}nchen}

\usepackage[dvipsnames,svgnames,x11names]{xcolor}
 \usepackage[final]{changes}

\definechangesauthor[color=BrickRed]{RR}

\usepackage{todonotes}
\setcommentmarkup{\todo[color={authorcolor!20},size=\scriptsize]{#3: #1}}

\newcommand{\note}[2][]{\added[#1,comment={#2}]{}}

\begin{document}

\author{R. Rizzato}\affiliation{\affA}
\author{F. Bruckmaier}\affiliation{\affA}
\author{K.S. Liu}\affiliation{\affA}
\author{S.J. Glaser}\affiliation{\affA}\affiliation{\affB}
\author{D.B. Bucher*}\affiliation{\affA}

\title{Polarization transfer from optically-pumped NV center ensembles to multinuclear spin baths}
\begin{abstract}
NV-diamonds have attracted keen interest for nanoscale sensing and spin manipulation. In particular, the non-equilibrium electron spin polarization after optical excitation of single NV centers has successfully been transferred to nuclear spin baths in the surrounding of the defects. However, these experiments need to be extended to NV-ensembles which have promising practical applications in the hyperpolarization of bulk sample volumes for NMR signal enhancement. Here, we use a dense, shallow ensemble of NV-centers to demonstrate polarization transfer to nuclear spins in a well-defined composite diamond sample system. This allows us to address three different types of nuclear spins in different positions with respect to the NV polarization source: from the close proximity of $^{13}$C inside the diamond lattice to the self-assembled molecular system consisting of $^{1}$H and $^{19}$F spins outside the diamond and over multiple interfaces. We show that ensemble NV experiments face problems different from single NV experiments. In particular, using spinlock pulses, \deleted[]{the T2* of the NV ensemble} \added[id=RR]{the inhomogeneously broadened ESR line of the NV ensemble} limits the minimal resonance linewidth with which the transfer protocol can occur. Furthermore, we compare the NV spin-polarization losses and polarization transfer rates to the different nuclear baths and discuss the role of spin-diffusion as detrimentally affecting the direct observation of nuclear polarization build-up within the detection volume of nanoscale NV-NMR experiments.
\end{abstract}

\maketitle 


\section*{Introduction}
\begin{figure*}[ht]
  \centering
  {\includegraphics[width=0.85\textwidth]{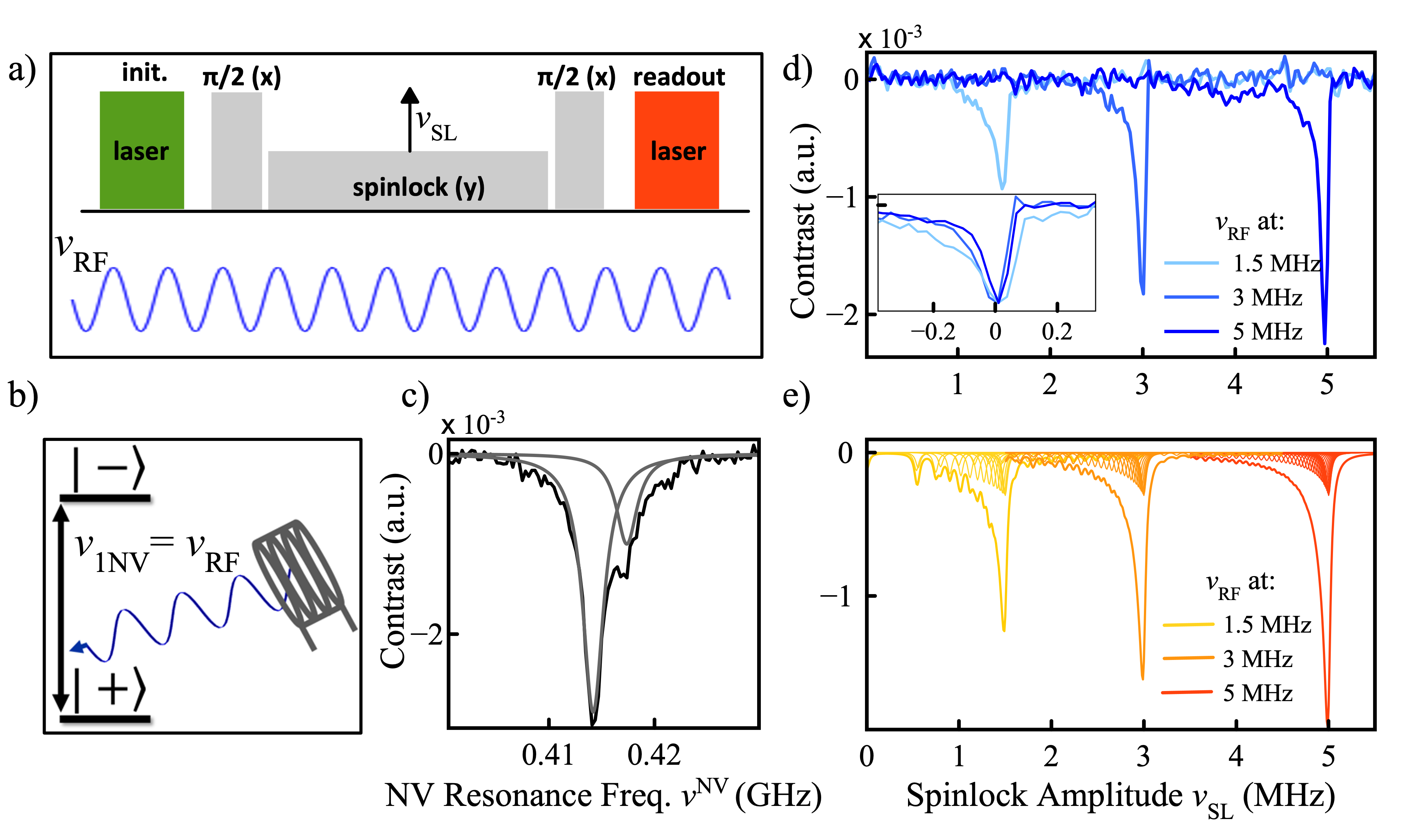}}
  \caption{\textbf{Spinlock experiment, calibration and characterization of the pulse sequence for an NV-ensemble.} a) Spinlock pulse sequence. First, a short laser pulse is applied to optically polarize the NV ensemble into the $m_{s}$=0 state. Then a MW pulse sequence consisting of $\frac{\pi}{2}_X$ – $spinlock_Y$ - $\frac{\pi}{2}_X$ pulses is applied. The final spin state is read out by fluorescence with a photodiode. Throughout these events, an oscillating magnetic field with frequency $\nu_{RF}$ in the MHz range is continuously applied with a wire loop (blue). b) Schematic of the experiment with matched radiofrequency. Transitions between electronic dressed states are driven when a coil generates a RF field that matches the splitting. c) ESR spectrum of the NV-ensemble corresponding to excitation of the NV transition from the $m_{s}$=0 to the -1 state. The spectrum is split by the $^{15}$N spin of the NV center and fit with 2 Lorentzian functions. d) Calibration of the spinlock's amplitude. The RF is kept at a fixed frequency and the spinlock pulse's amplitude is swept. NV depolarization appears as asymmetric dips in the optical contrast as soon the MW field strength matches the RF frequency. Inset: the three dips are overlayed for a linewidth comparison. e) Simulated depolarization dips (see Appendix 2 for details).}
 \zfl{fig1}	
\end{figure*}
Nuclear Magnetic Resonance (NMR) encompasses a wide range of techniques which are of fundamental importance for the chemical characterization of organic and inorganic materials as well as of biological and complex systems. However, NMR spectroscopy lacks sensitivity, due to the extremely low nuclear spin polarization that generates the detected signal. To overcome the issue, hyperpolarization methods, such as Dynamic Nuclear Polarization (DNP), have emerged consisting of well-suited microwave (MW) irradiation schemes that allow to transfer the higher thermal polarization of the electron spins to the target nuclei\cite{overhauser_polarization_1953,nelson_metabolic_2013,ardenkjaer-larsen_facing_2015,lilly_thankamony_dynamic_2017}. In contrast, the electronic spin of the NV-center in diamond can be highly polarized by optical excitation at room temperature and coherently manipulated, forming a promising hyperpolarization platform \cite{maze_nanoscale_2008,balasubramanian_nanoscale_2008,belthangady_dressed-state_2013,tetienne_prospects_2021}. Although direct detection of spin polarization build-up has not been demonstrated yet \added[id=RR]{for nuclei external to the diamond}, the polarization transfer from NV centers to nuclear spins has been accomplished either within the (nano)diamond lattice \cite{PhysRevB.81.073201,london_detecting_2013,PhysRevB.97.024422,Ajoyeaar5492,schwartz_robust_2018} or from single NV centers to small external nuclear ensembles. \cite{fernandez-acebal_toward_2018,shagieva_microwave-assisted_2018,broadway_quantum_2018}. However, the latter systems are limited to the nanoscale and are not suited for applications in micronscale NV-NMR experiments \cite{glenn_high-resolution_2018,smits_two-dimensional_2019,bucher_hyperpolarization-enhanced_2020,PRXQuantum.2.010305}, bulk NMR or biomedical imaging, where the ratio between number of NV defects and target nuclei needs to be significant. Scaling up the NV system by means of ensembles is then necessary, but not trivial since they face different problems due to their density and spatial distribution. Inhomogeneous properties throughout the ensemble, shorter coherence times and inhomogeneities of both the driving MW field and the static magnetic field are only examples of the issues that need to be tackled when working with NV-ensembles and that are to be investigated in detail. 

Here, we report on the use of shallow-implanted NV-ensembles for the direct transfer of optically pumped electron spin-polarization to nuclear spins internal to the diamond lattice and external to a well-defined Self Assembled Monolayer (SAM) \cite{liu_surface_2021}. Such a system has been chosen since it has several advantages: 1) it is a well defined solid state sample with known surface chemistry, 2) it contains several types of NMR active nuclei that can be addressed at the same time, 3) it permits the exploration of spin-polarization transfer over multiple interfaces and 4) its thin structure of 2-3 nanometers allows us to exploit a certain degree of confinement for external nuclear polarization accumulation and storage. 

The polarization transfer is accomplished through an experiment known as Nuclear Spin Orientation via Electron Spin Locking (NOVEL) which belongs to the family of the NMR signal enhancement techniques, more specifically pulsed \deleted[id=RR]{Dynamic Nuclear Polarization} DNP experiments \cite{henstra_nuclear_1988,brunner_notizen_1987,scheuer_dynamic_2020}. Here, the electronic spin of the NV defect in diamond is brought into a superposition state by a strong $\frac{\pi}{2}$ pulse and then locked on the transversal plane of the laboratory frame by a long spinlock pulse of variable amplitude. In this manner, an electron spin dressed state is created and preserved for a time interval on the order of its rotating frame spin-lattice relaxation time $T_{1\rho}$, namely few hundreds of microseconds. During this time, the electronic dressed state can be brought in energy contact with the nuclear spin by matching the MW field strength used to lock (or ``dress'') the electronic spin with the Larmor frequency of the target nuclear spins ($\nu_{1NV}=\nu_{0n}$). This is the so-called \textit{Hartmann-Hahn} (HH) matching condition \cite{PhysRev.128.2042}, well-known in NMR for conducting cross-polarization experiments between different nuclear spins but also, in some of its variants, in Electron Spin Resonance (ESR) and DNP to accomplish polarization transfers between electron and nuclear spins \cite{weis_electron-nuclear_2006, rizzato_cross-polarization_2015,scheuer_robust_2017,scheuer_dynamic_2020}. 

In order to probe the actual polarization of the nuclear target, we compare the effects observed with the NOVEL experiment with that resulting from the Polarization Readout via Polarization Inversion (PROPI) experiment \cite{,scheuer_dynamic_2020,schwartz_robust_2018,shagieva_nmr_2019}. In PROPI, two NOVEL pulse sequences are applied one after the other, differing only in the spinlock pulse phase which is set to polarize the nuclear spins in two opposite directions. Then, if the first block of the sequence polarizes the nuclear spins in the ``up" state, the second polarizes in the ``down'' state or vice versa. In this manner, it is possible to probe the extent to which the nuclear ensemble has been effectively polarized and preserves the polarization throughout the experimental timescale. We explore the efficiency of the spinlock pulse sequence and its potential to build-up nuclear polarization by cycling the experiment in a competition with nuclear relaxation and spin diffusion processes.
\section*{Results}
\textbf{Characterization of the spinlock experiment with NV ensembles.} Before performing the actual experiments on nuclear spin samples, it is useful to characterize the pulse sequence by means of an external radiowave frequency (RF) field in the MHz range that mimics the oscillating magnetic fields generated by nuclear spin precession. Since such an experiment does not rely on actual polarization transfer to nuclear spins, we will distinguish it from NOVEL by simply calling it ``spinlock" experiment. For this purpose, an RF coil has been mounted few millimeters above the diamond chip in the experimental setup. The coil has been oriented such to generate an oscillating magnetic field parallel to the orientation of the external $B_0$ field. The spinlock pulse sequence is set up as shown in \zfr{fig1} a). Driving the NV ensemble at the resonance frequency and sweeping the spinlock field amplitude while continuously irradiating at a certain radiofrequency results in a fluorescence contrast dip as soon the NV Rabi frequency matches the frequency of the applied RF field. This can be described considering the NV spins to be in a dressed states basis, where the new quantization axis lies on the x-y plane of the laboratory frame. Matching the energy splitting by means of an oscillating field along the z-axis, can induce transitions between the dressed ``plus'' state and the ``minus" state (or vice versa), leading to a reduction in fluorescence (see \zfr{fig1} b)). In \zfr{fig1} d) the experiment is shown for RF reference signals with different frequencies. From this it is possible to build a calibration curve of the applied RF frequency versus the spinlock MW amplitude as shown in Appendix 3, \zfr{figb1}. In contrast to single NV-experiments, the depolarization dips show a characteristic lineshape which is tailed towards lower MW amplitudes and becomes overall sharper when higher reference frequencies are matched. Such a behavior can be explained by the fact that during the experiment, the RF field is set at a constant frequency and is matched with the Rabi frequencies of the NV ensemble: $\nu_{RF}=\nu_{1NV}$. Distinct from the single NV case, in a dense ensemble, the ESR resonance line (reported in \zfr{fig1} c)) is inhomogeneously broadened, leading to a linewidth of $\sim$2.4 MHz in our case. Although we can assume a full excitation of the ensemble since the MW excitation bandwidth covers the entire ESR spectrum (as the former in our case is around 10 times larger), the various ``spin packets'' (homogeneously broadened components of the spectrum, see Appendix 2 and \zfr{linewidth}) will resonate at frequencies $\nu_i^{NV}$ slightly different from the NV resonance frequency $\nu_0^{NV}$ at which we tune our MW (normally the center of the ESR line). Thus, although with different contributions, they will all take part in the spinlock process but the NV Rabi frequency $\nu_{1NV}$ will rather be a distribution of ``effective" driving fields, such as: $\nu_{1NV,i}=\sqrt{{\Omega_i}^2+{\nu_{SL}^2}}$, where $\Omega_i=\nu_0^{NV}-\nu_i^{NV}$ is the offset frequency for the $ith$ spin packet (or detuning) and $\nu_{SL}$ the MW amplitude of the spinlock pulse. For instance, for the $0th$ spin packet at \textit{exact resonance}, $|\Omega_0|=0$ and the NV Rabi frequency would purely correspond to the spinlock amplitude that is matched to the RF: $\nu_{1NV,0}=\nu_{SL}=\nu_{RF}$. Whereas, for off-resonant NV spin packets, $|\Omega_i|\neq 0$ and this causes the Rabi frequency to be shifted from the spinlock amplitude which can no longer satisfy the matching condition, being: $\nu_{1NV,i}> \nu_{SL}=\nu_{RF}$. Thus, since $\nu_{RF}$ is kept constant, a smaller value of $\nu_{SL}$ will be required to match the condition with $\nu_{1NV,i}$ and these off-resonant NV centers will all contribute to provide the observed asymmetry as also described in Appendix 2. Moreover, the dip becomes sharper at high frequencies since the off-resonant contributions become less and less important with respect to the driving field $\nu_{SL}$ applied. Simulation of such features have been reported in \zfr{fig1} e) and have been computed considering the contributions to the depolarization lineshape arising from the off-resonant spin packets of the NV ensembles within the ESR spectral linewidth, as detailed in Appendix 2. As we will see later, such spectral features are preserved during the NOVEL/PROPI experiments on actual nuclear spins. As a result we conclude that higher matching frequencies are preferable using NV-ensembles. Another effect influencing the depolarization dip linewidth and the overall performance of the spinlock experiments, arises from the inhomogeneity of the driving MW field over the NV ensemble (see Appendix 1).   

\textbf{A composite multinuclear system for polarization transfer.} As a model system, the surface of a diamond chip \added[id=RR]{previously implanted with $\sim$5 nm deep NV centers} is functionalized with a 1 nm layer of Al$_2$O$_3$ prepared by Atomic Layer Deposition (ALD). On such a thin film, a Self Assembled Monolayer (SAM) of polyfluorinated organic molecules has been grown\cite{liu_surface_2021} (see \zfr{fig2}). With this system we can address three types of nuclear spins that are in very different conditions with respect to their interaction with the NV ensembles. i) $^{13}$C nuclei in natural abundance are located inside the diamond lattice, in close proximity to the NV centers and interact strongly; ii) $^{1}$H spins are known to be ubiquitous at the diamond surface\added[id=RR]{\cite{staudacher_probing_2015}}, though their properties (density, spatial distribution and location, mobility, physical state etc.) are unknown; iii) $^{19}$F nuclei are bound to the molecules constituting the SAM, their spin has a natural abundance of 100 percent and a gyromagnetic ratio comparable to that of protons, but unlike from $^{1}$H, they can be unambiguously assigned to our well-defined surface \cite{liu_surface_2021}. It is important to keep in mind that our NV-ensemble is a rather dilute system, where we estimate an average distance of NV centers $\sim$100 nm from each other\cite{ziemQuantitativeNanoscaleMRI2019, devienceNanoscaleNMRSpectroscopy2015}. Every NV can be then considered as an independent spin system throughout the sample space dictated by our laser spot size (around 4000 $\mu $m$^2$). For this reason, all effects observed throughout this work will be referred to a system composed by a single NV center and an ensemble of nuclear spins.

\begin{figure}[t]
\centering
{\includegraphics[width=0.5\textwidth]{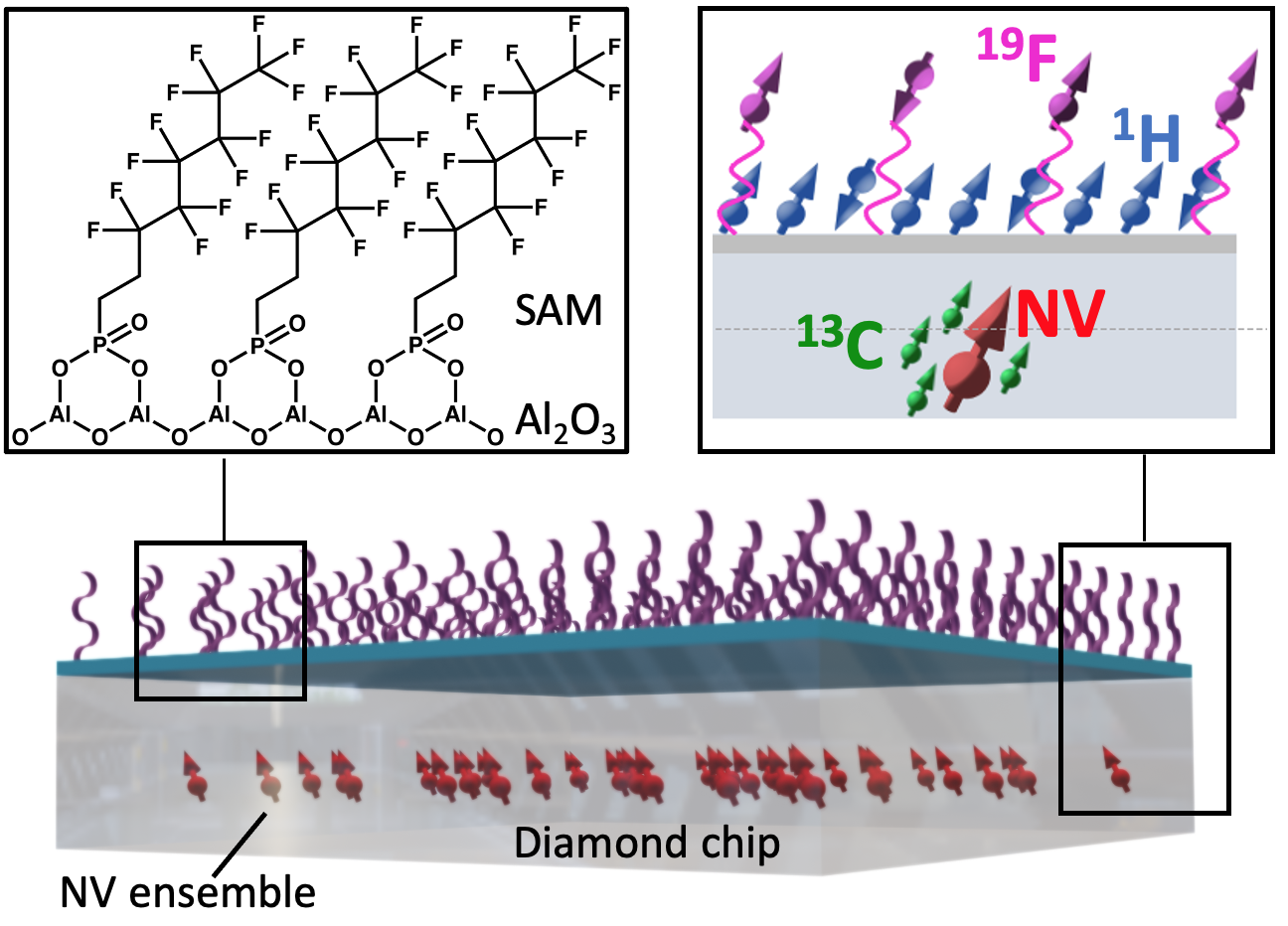}}
\caption{\textbf{A composite multi-nuclear system for polarization transfer}. Schematic of the probed diamond system. A chip containing ensembles of near-surface NV-centers has been ALD coated with a thin (1 nm) film of Al$_2$O$_3$ modified with a monolayer of 1H, 1H, 2H, 2H-perfluoroctanephosphonic acid (PFOPA) molecules. The NV ensemble is used to transfer spin polarization to the nearby $^{13}$C spins and to $^{1}$H and $^{19}$F spins outside of the diamond lattice.}
\zfl{fig2}	
\end{figure}

\textbf{Polarization transfer to nuclear spins.}   
As schematically depicted in \zfr{fig3} a), during a NOVEL experiment, the NV dressed states are split by their Rabi frequency $\nu_{1NV}$ and can be brought in energy contact with the nuclear bare states by matching the MW amplitude with the nuclear Larmor frequency. Such a protocol can be repeated many times, since the NV spin can be reinitialized in the $m_s$ = 0 state by laser excitation at each time. In each of these repetitions, spin polarization transfer can occur and a consequent drop of the NV spin signal is detected as a decrease in photoluminescence (PL) (depolarization dips). However, if during a NOVEL experiment the interacting nuclear bath becomes completely polarized, a ``saturation'' of the system would cause the depolarization dips to be no longer observable. In a PROPI experiment (\zfr{fig3} b)), two NOVEL sequences are run consecutively,  differing only in the phase of the spinlock pulses which is shifted by $\pi$ with respect to each other \cite{scheuer_robust_2017}. In this manner, the second spinlock pulse neutralizes the nuclear polarization generated by the first one allowing the dips to be always observable, independently on the nuclear spin bath‘s polarization level.
A comparison of the NOVEL (red line) versus PROPI (blue line) is shown for our composite multi-nuclear system. The amplitude of \added[id=RR]{a 40 $\mu$s} spinlock pulse is swept in a range corresponding to Larmor frequencies of the probed nuclear spin types at a magnetic field $B_0$ of 88 mT. The PROPI sequence reveals four intense depolarization dips, where the signal at 2.5 MHz corresponds to a RF reference signal and the dips around 0.94, 3.55 and 3.73 MHz correspond to matching with the Larmor frequencies of $^{13}$C, $^{19}$F and $^{1}$H nuclei, respectively(\zfr{fig3} c), blue line). Because of the similar  gyromagnetic ratios of $^{19}$F and $^{1}$H, the Larmor frequencies differ by less than 200 kHz and the signals are partially overlapped. Nevertheless, as shown in \zfr{fig3} d) they can be clearly distinguished. 
A different outcome is reported when using the NOVEL pulse sequence (\zfr{fig3} c), red line). In this case, the $^{13}$C dip disappears \added[id=RR]{and even undergoes an inversion}, whereas all the other peaks, also the reference one at 2.5 MHz, preserve their lineshapes and amplitudes.\added[id=RR]{The experimental conditions for both experiments are kept the same, each spectral point being the result of  5000 repetitions of the pulse sequence (around 100 $\mu$s long) and 20 averages of the full amplitude sweep for an experimental time of around 100 s per sweep. The signal has been then computed as the difference of the two consecutive optical readouts (\zfr{fig3} b)) divided by their sum and then normalized to the intensity of the reference signal.}  Note that peaks at higher frequencies are narrower and that NOVEL/PROPI experiments result in the same lineshapes as in the reference spinlock experiments described in \zfr{fig1} e). Simulations showing the effect of the off-resonant NV defects to the overall lineshape have been performed and reported in \zfr{fig3} e). Finally, the same experiments have been performed on a clean diamond upon removal of the SAM and Al$_2$O$_3$ layer by means of acid treatment (yellow line, \zfr{fig3} d)). As expected, the $^{19}$F signal now is absent and only signals due to $^{13}$C and $^{1}$H matching can be observed (see full sweep spectrum in Appendix 3).

\begin{figure}[t]
\centering
{\includegraphics[width=0.5\textwidth]{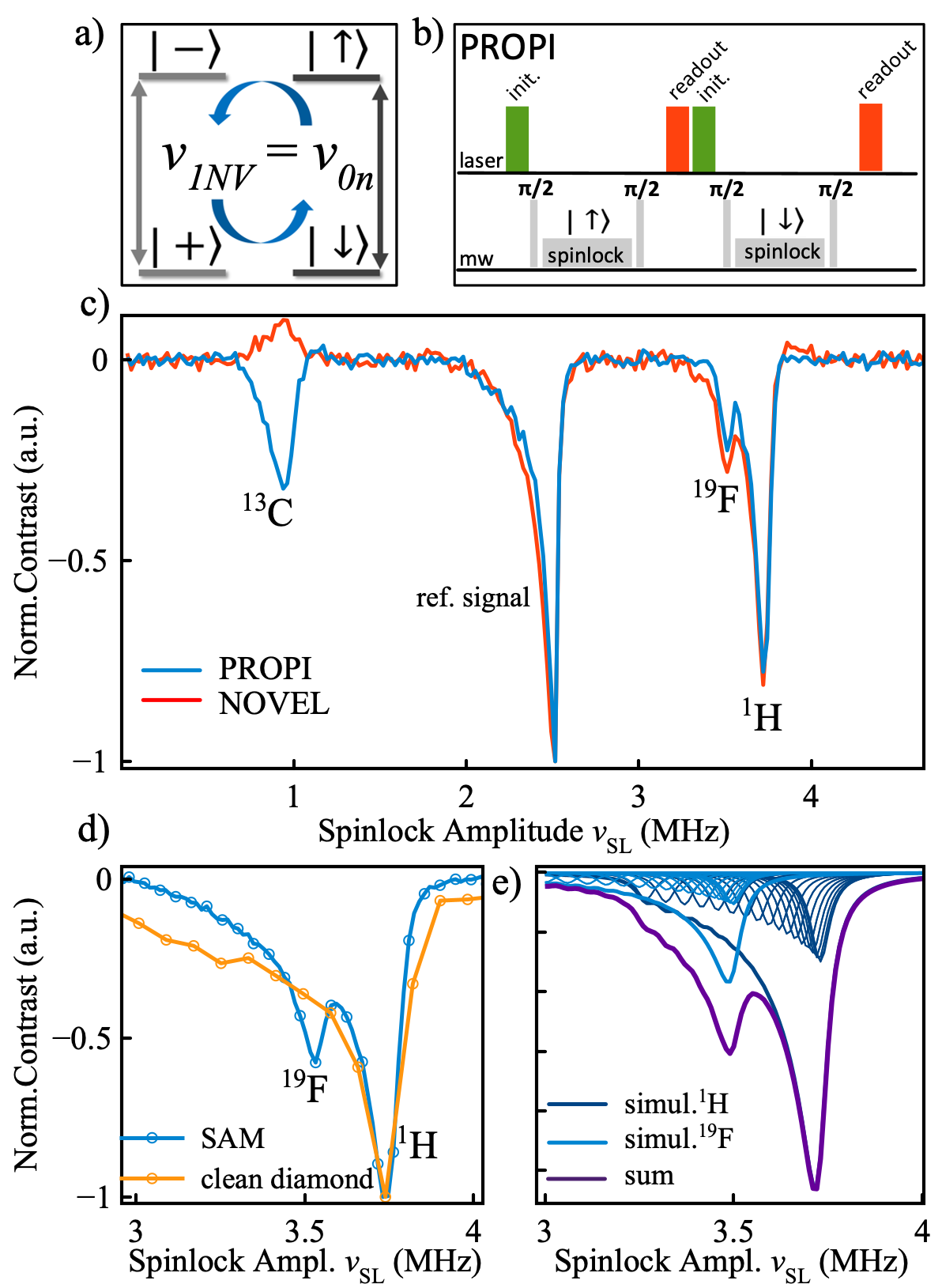}}
\caption{\textbf{Nuclear spin detection by matched Hartmann-Hahn condition with an NV-ensemble}. a) Schematic of the energy matching mechanism between electron spin dressed states and nuclear spin bare state. b) PROPI pulse sequence. c) PROPI versus NOVEL experiment. Matched HH condition corresponding to $^{13}$C, $^{19}$F and $^{1}$H resonances results in a depolarization of the NV-spin states and consequent dips in PL contrast for PROPI, blue line, whereas the $^{13}$C dip disappears in the NOVEL experiment (red line). A dip from a reference signal of 2.5 MHz frequency is also shown. d) Detailed PROPI depolarization spectrum around the $^{19}$F and $^{1}$H Larmor frequencies. The line in yellow shows the same spectrum for a cleaned diamond without the surface modification. e) Simulated lineshape, see Appendix 2.} 
\zfl{fig3}	
\end{figure}

\textbf{Study of the polarization transfer dynamics.} In the next step we extract dynamic information about the transfer process, by monitoring the fluorescence contrast while increasing the spinlock pulse duration (\zfr{fig4}). As the spinlock pulse gets longer, the detected signal evolves in an exponential decay which is mainly determined by two simultaneous processes. First, electron spin relaxation will manifest with a rotation-frame characteristic time $T_{1\rho}$. This parameter can be easily measured by setting the spinlock amplitude such that the HH condition is not satisfied. This corresponds to the ``unmatched HH'' curves in \zfr{fig4} a), c) and e) whose fittings gave $T_{1\rho}$ time constants of $\sim$109 $\mu$s for $^{13}$C, and $\sim$225 $\mu$s for $^{19}$F and $^{1}$H.  In the second case, at the HH condition, enhanced signal loss is caused by additional matching with the nuclear spin bath, i.e. polarization loss to the spin baths (see \zfr{fig4} a), c) and e). We assume that, for each spinlock duration used in the experiment, such an excess of depolarization observed in the matched case versus the non-matched one is due to spin polarization transfer to the nuclear target. Thus, the contribution due to relaxation mechanisms can be distinguished from the polarization transfer by simply subtracting the two curves (\cite{tetienne_prospects_2021}) (\zfr{fig4} b), d) and f)). The maximum of such difference curves would then correspond to an experimental depolarization of the NV spin of around 4.5 $\%$ after 5 $\mu$s spinlock for $^{13}$C, 1.5 \% after 40 $\mu$s for $^{1}$H and 1 \% after 65 $\mu$s for $^{19}$F. 
\begin{figure}[t]
\centering
{\includegraphics[width=0.5\textwidth]{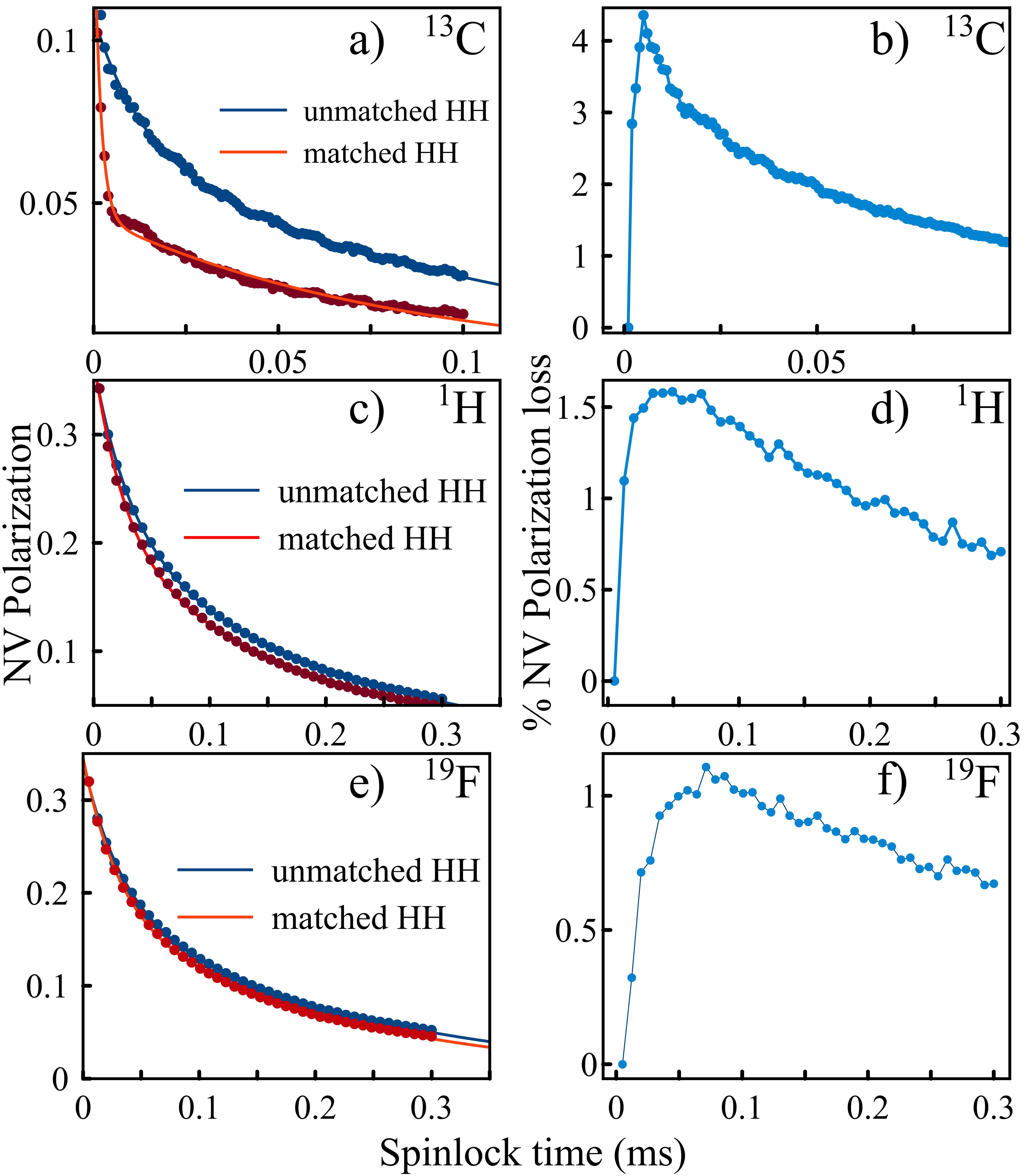}}
\caption{\textbf{Polarization transfer dynamics.} NV spin polarization during PROPI experiment is monitored while varying the spinlock time in the case of matched (red dots) and unmatched (blue dots) HH condition for $^{13}$C a) $^{1}$H c) and $^{19}$F e). By subtracting the matched and unmatched data points, respective difference curves in b), d) and f) are obtained showing the percentage of NV-spin depolarization due to polarization transfer.
The decays were fit with a biexponential function as reported in Appendix 1.}
\zfl{fig4}	
\end{figure}
\begin{figure}[t]
{\includegraphics[width=0.45\textwidth]{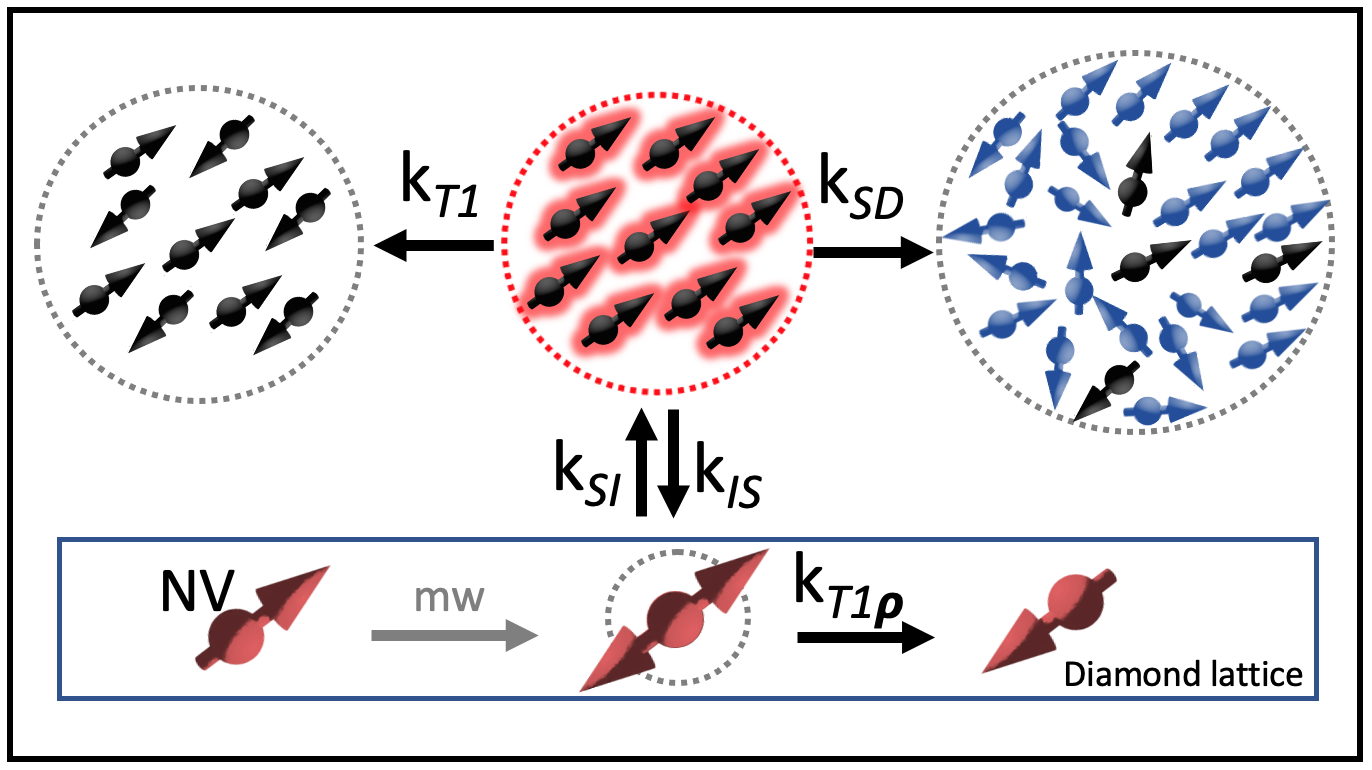}}
\caption{\textbf{Simplified rate equation model for polarization built-up.} Schematic representation of a NV ensemble brought in thermal contact with a bath of target nuclear spins. The rate constants of polarization exchange between NV and nuclear ensemble are represented by $k_{SI}$=$k_{IS}$, the one for polarization loss by spin diffusion to other nuclei is $k_{SD}$ and the nuclear spin-lattice relaxation rate $k_{T1}$.}
\zfl{fig5}	
\end{figure}
\begin{figure*}[ht]
{\includegraphics[width=0.75\textwidth]{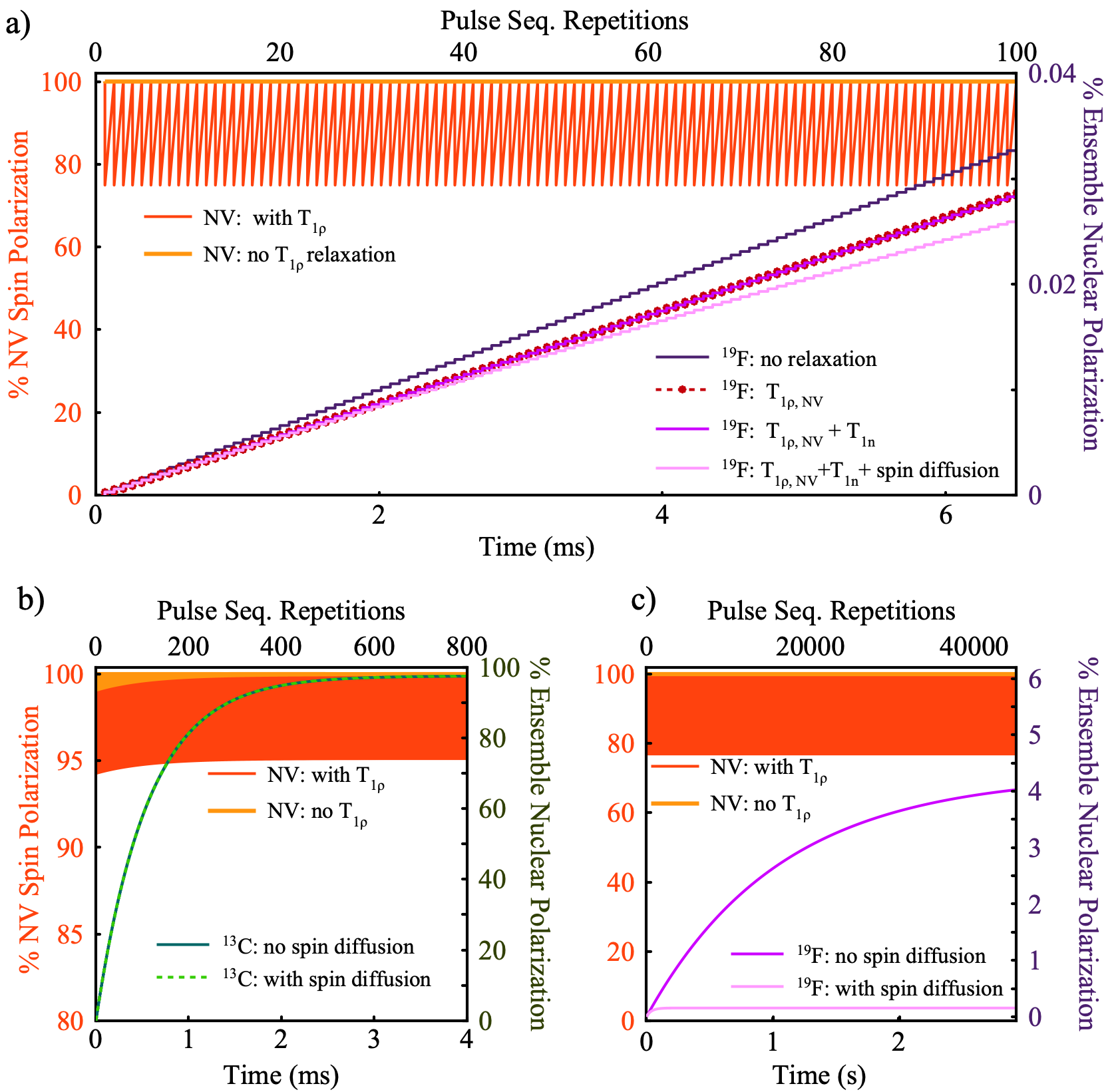}}
\caption{\textbf{Simulated polarization build-up curves.} Simulation of the NV and nuclear spin polarisation versus the number of repetitions of the pulse sequence (or the total experimental time) considering the transfer between one NV and the nuclear spin ensembles. a) First 100 repetitions for the transfer from one NV to an ensemble of 3000 $^{19}$F nuclear spins. Each step corresponds to a 65 $\mu$s polarization transfer protocol considering the effective rate determined above for this spin (154 s$^{-1}$/3000). The red and yellow lines represent electron spin polarization with and without NV relaxation ($T_{1\rho}$), respectively. The lines from violet to pink show how nuclear buildups are affected by the different ``loss channels'' considered in the system. b) Transfer to an ensemble of five $^{13}$C spins occurring upon 800 repetitions of the sequence. Steps are 5 $\mu$s long at the effective rate determined for this spin (9000 s$^{-1}$/5). c) Transfer to the $^{19}$F spin bath up to 45000 repetitions. Both $T_{1n}$ and $T_{1\rho}$ have been considered for the nuclear buildups in b) and c).}
\zfl{fig6}	
\end{figure*}
\section*{Discussion}
Our observations show that spin polarization of the NV undergoes significant losses upon application of spinlock-based pulsed schemes at the HH matching condition. A polarization transfer to all three nuclear species addressed in this work can thus be anticipated. 
In particular, transfer to $^{13}$C spins inside the diamond is expected to be an efficient process due to the close proximity of the nuclei to the NV and the consequent strong couplings (up to hundreds of MHz) which leads to transfer rates several orders of magnitude faster than nuclear spin relaxation rates \cite{PhysRevB.79.075203,Smeltzer_2011,PhysRevB.85.134107,alvarez_local_2015}. The depolarization dip in \zfr{fig3} c), gets indeed fully suppressed by the NOVEL experiment, indicating that total polarization of the target nuclear ensemble has been achieved. Moreover, an inversion of the detected contrast is observed. \deleted[id=RR]{This is probably due to the electron spin $T_1$ relaxation that, once full polarization of the $^{13}$C spin bath is reached, constitutes a preferential pathway to re-equilibrate the system by repopulating the NV $m_s$ = 0 state during the spinlock time.} \added[id=RR]{This is possible since after a certain number of pulse sequence repetitions, the $T_{1\rho}$ relaxation acting during the spinlock, causes the NV spin polarization to decay at a polarization level which is lower than what the $^{13}$C polarization buildup has been able to reach. Thus, a polarization transfer from the nuclear ensemble back to the NV ensemble can occur, as described in detail in Appendix 4.}
In the experiments shown in \zfr{fig4} a) and b), the depolarization of the NV spin reaches about 4.5 \% after 5 $\mu$s spinlock time. Thus, assuming that an interacting $^{13}$C spin bath would symmetrically gain spin polarization and that it will be shared by all nuclear spins in the ensemble (proportionally to their coupling to the NV), a polarization of 4.5 \%/$N$ is expected for the whole bath, where $N$ is the number of the nuclei populating it. This would result in a rate constant for such a process of $k_{SI} $=\;9000 $s^{-1}$, resulting in a timescale on the order of 2-3 ms to polarize a few $^{13}$C nuclei in the close surrounding of an NV center. 
On the contrary, in the case of nuclei more distant from the NV centers, such as $^{1}$H and $^{19}$F spins outside the diamond, the weaker interactions determine much slower rates as shown in \zfr{fig4} c) and e). Moreover, longer times would be needed for a complete polarization of these ensembles, since the number of spins is higher compared to the $^{13}$C bath. The effective polarization transfer rates measured in the experiments in \zfr{fig4} d) and f) amount to 375 s$^{-1}$ and 154 s$^{-1}$ for $^{1}$H and $^{19}$F, respectively. This means for instance that build-up times on the order of several seconds for ensemble of few thousand spins are expected  \note[id=RR]{earlier in the text?}($\sim$3000 $^{19}$F spins are estimated to be present in our monolayer in the proximity of an NV center).
\added[id=RR]{It is worth to note that our simulations consider an ideal NV spin polarization of 100$\%$. However, as reported in \zfr{fig4}, the actual polarization available for the transfer during our experiments was only a fraction of it, i.e., about 10$\%$ for the experiments on $^{13}$C and 30$\%$ for the $^{1}$H and $^{19}$F matchings. Thus, the maximum nuclear polarization achievable in these conditions scales accordingly. Such a behaviour is likely due to the low MW amplitude of the driving field which causes inefficient spin locking and consequent fast NV T$_{1\rho}$ relaxation. Such an issue motivates to work at higher fields where the matching fields will be proportionally higher in amplitude resulting in more efficient spinlock pulses.} Nevertheless, considering that these dynamics occur on the same time-scale as nuclear spin lattice relaxation $T_{1n}$,  nuclear polarization build-up is still expected to be observable, at least to some extent. However, no effect has been observed in the NOVEL experiments described above for such nuclei. The reason for this can also be found in spin diffusion that spreads the polarization over a larger volume by means of the nuclear dipolar coupling network. In conventional bulk DNP, spin diffusion is important since it contributes positively to nuclear polarization build-up. However, under our nanoscale conditions the polarization pumped from the NV would be rapidly lost out of the sensing volume so that its accumulation becomes hardly observable with our experiment. 
In order to rationalize the experimental findings, a simplified kinetic model is set up to describe, at least to a first approximation, the expected build-up of nuclear polarization by continuous repetition of the NOVEL pulse sequence in the presence of simultaneous ``loss channels'' (see \zfr{fig5}). Here, we consider the transfer of spin polarization to take place together with three interfering processes:  electron spin relaxation in the rotating frame ($T_{1\rho}$), nuclear spin-lattice relaxation ($T_{1n}$) and spin diffusion ($T_{SD}$). To each of these, specific rate constants are associated and in particular: $k_{SI}$ = $k_{IS}$ and $k_{T1\rho}$ are the rates for the polarization transfer and the NV $T_{1\rho}$ relaxation, respectively and have been measured experimentally as reported above. The nuclear relaxation $k_{T_1n}$ and the spin diffusion $k_{SD}$ rates are both estimated based on data reported in the literature\cite{broadway_quantum_2018,ajoy_hyperpolarized_2019,tetienne_prospects_2021} (see Appendix 4). Simulations have been performed calculating for each repetition of the sequence the change of the nuclear and electron spin polarizations obtained by the solution of the system of equations reported in Appendix 4. In \zfr{fig6} a), the simulation of both the NV and nuclear polarizations during the first 100 repetitions of the pulse sequence is depicted for the case of polarization transfer to a $^{19}$F spin ensemble. The NV polarization is initialized at each repetition, differently from the nuclear polarization which is built up over time. First, it can be noted that the $T_{1\rho}$ relaxation process has a significant effect on the depolarization of the NV (yellow and red lines) but a rather small on the nuclear polarization (violet and dotted magenta line), even at longer timescales. This parameter does not really represent a limit for reaching appreciable polarization buildups. In \zfr{fig6} b), the buildup curves for $^{13}$C are simulated showing around 800 repetitions of the experiment on a timescale on the order of few milliseconds to fully polarize the $^{13}$C spin bath. In this case both spin diffusion and nuclear spin-lattice relaxation are negligible, since they occur on a time scale much longer than the polarization transfer process. 
On the contrary, for $^{19}$F spins, spin diffusion seems to have a dramatic effect on the polarization build-up causing the steady state polarization to decrease by more than an order of magnitude with respect to what is expected when only spin-lattice relaxation is considered (see \zfr{fig6} c)). Such a hypothesis appears realistic especially if we think that heteronuclear spin diffusion between $^{19}$F and $^{1}$H may also play a role at the low magnetic field at which experiments have been performed.
Thus, to accomplish a direct NV-based detection of enhanced NMR signals, two main approaches can be envisioned: 1) increase the polarization transfer rates; 2) polarization loss pathways need to be circumvented or at least significantly mitigated. About the first point, \added[id=RR]{performing spinlock experiments at higher magnetic fields might be of significant help for the possibility to match the higher nuclear Larmor frequencies with consequently higher spinlock amplitudes. This would  advantage the efficiency of the spinlock pulse and of the overall transfer process at least by a factor 3. Moreover, the use of a resonator for mw delivery instead of the homebuilt mw loop as employed in the present work would reduce significantly the MW field inhomogeneity and greatly improve the performance.} \added[id=RR]{Another} possible strategy would be to explore alternative polarization transfer schemes which are more robust in particular for NV-ensembles \added[id=RR]{and would allow to improve the transfer rates by more than one order of magnitude} \cite{schwartz_robust_2018, healey_hyperpolarisation_2021}. However, these schemes are better performed on isotopically enriched $^{12}$C diamonds since they can be strongly affected by spurious harmonic signals caused by nearby $^{13}$C nuclei \cite{PhysRevX.5.021009}.  About the second approach, an appropriate engineering of the sample will allow to conveniently tune the properties of the system such to decrease diffusion and relaxation. For instance, creating a more ``diluted'' nuclear spin system with reduced nuclear spin-spin interactions as well as, working at higher magnetic fields to inhibit heteronuclear spin diffusion will avoid polarization loss out of the NV-detection volume. 

\section*{Conclusion}
In summary, we demonstrate spin-polarization transfer between ensembles of shallow NV-centers in diamond and nuclear targets inside and outside the diamond lattice. We found that spinlock pulse sequences based on Hartmann-Hahn matching, such as NOVEL, for the special case of the ensembles have characteristic spectral features. These have to be taken into account and suggest to work preferably with higher matching frequencies (i.e. at higher magnetic fields). Furthermore, when performed to address actual nuclear spins, such protocols can lead to significant depletion  of NV spin polarization by HH matching with nuclear spins, even when nuclei are outside the diamond and behind multiple material interfaces. The chemical functionalization of our diamond chip was a well-defined system where several types of nuclear spins could be addressed and their results compared. In the case of nuclei far from the NV centers, the polarization transfer rates need to be increased to overcome some counteracting processes causing nuclear polarization losses. Apart from nuclear spin relaxation, we identified spin diffusion as the additional process which may hinder a direct NV-based detection of the transferred polarization. A simplified model based on rate constants has been used to estimate the contribution of spin diffusion to the maximum polarization achievable when typical nuclear spin lattice relaxation times are considered. This allows us to better define future strategies that can be pursed to achieve the final goal. Based on these results, the implementation of platforms where diamond surface-chemistry can be controlled with high accuracy and flexibility, confirm to be key to establish NV-based methods for NMR sensitivity enhancement.

\textbf{Acknowledgements.}
Funding: This study was funded by the Deutsche Forschungsgemeinschaft (DFG, German Research Foundation) - 412351169 within the Emmy Noether program. R.R. acknowledges support from the DFG Walter Benjamin Programme (project RI 3319/1-1). S.G. and D.B.B acknowledges support from the Deutsche Forschungsgemeinschaft (DFG, German Research Foundation) under Germany’s Excellence Strategy – EXC-2111 – 390814868.
\textbf{Author contributions.}
D.B.B. and R.R. conceived the idea, R.R. carried out the experiments, analyzed the data and performed the simulations, F.B. programmed the pulse sequences, K.S.L. prepared the ALD sample, S.J.G. advised on the theoretical aspects of the work. R.R. wrote the manuscript with inputs from all authors. All authors reviewed the manuscript and suggested improvements. D.B.B. supervised the overall research work.
\textbf{Competing interests.}  All authors declare that they have no competing interests. \textbf{Data and materials availability.} All data needed to evaluate the conclusions in the paper are present in the paper and/or the Supplementary Materials. Additional data related to this paper may be requested from the authors. All correspondence and request for materials should be
addressed to D.B.B. (dominik.bucher@tum.de).

\textit{Note added.} \added[id=RR]{During the preparation of the manuscript we have become aware of the publication from Healey \textit{et al.}\cite{healey_hyperpolarisation_2021} also demonstrating polarization transfer by means of NV ensembles in diamond. In relation to this work, our study is different and complementary for the following reasons$:$ $1)$ The focus of this work is to implement the spinlock experiments for NV ensembles and understand the spectral features. The polarization transfer in our work relies on using NV dressed states generated by spinlock pulses, a different approach with respect to  Healey \textit{et al.} (Pulsepol). $2)$ Our chemical system is totally different, since we address monolayers deposited on the diamond surface  and explore the potential of such quasi-2D structure for nanoscale spin hyperpolarization. $3)$ For the first time, direct polarization transfer has been achieved to $^{19}$F nuclei unambiguously positioned outside the diamond lattice and across an additional interface (Al$_2$O$_3$ thin film). $4)$ With this study we propose the use of chemical structures grown on the diamond surface  with well-defined characteristics as a testbed for NV-based spin hyperpolarization at the nanoscale.}
\section*{Appendix 1: Materials and Methods}
\textbf{Diamond properties and sample preparation.}
All experiments are based on a 2 x 2 x 0.5 mm electronic grade diamond (natural (1.1$\%$) $^{13}$C abundance, Element Six) which was implanted using $^{15}$N at an energy of 2.5 keV with a fluence of 2x10$^{12}$/cm$^{2}$ (Innovion) and subsequently annealed\cite{bucherQuantumDiamondSpectrometer2019b}. This results in a distribution of near-surface NV centers $\sim$5 nm below the surface. For these implant conditions, an NV density of $\sim$ 50-100 NVs/$\mu$m$^{2}$ can be estimated\cite{ziemQuantitativeNanoscaleMRI2019, devienceNanoscaleNMRSpectroscopy2015}. 
The T$_1$ and T$_2$ of our shallow NV-ensemble are $\sim$650 $\mu$s and $\sim$4 $\mu$s, respectively. Before performing experiments on clean diamond as well as before every Al$_2$O$_3$ deposition the diamond was cleaned with a protocol involving equal parts of boiling sulfuric, nitric, and perchloric acid (tri-acid cleaning) according to Brown \textit{et al.}  \cite{BROWN201940}. For the preparation of the system as shown in \zfr{fig2}, Atomic Layer Deposition (ALD) was performed for the deposition of a 1 nm thin Al$_2$O$_3$ film, followed by phosphonic acid surface functionalization and formation of monolayer as described in detail in Liu \textit{et al.} \cite{liu_surface_2021}. The number of spins considered throughout the work for the $^{19}$F ensemble (around 3000) has been estimated from the work of Liu \textit{et al.} for a similar system.

\textbf{Experimental setup.}
The experiment is based on a modified version of the setup described in Bucher \textit{et al.} \cite{bucherQuantumDiamondSpectrometer2019b}. The diamond chip was positioned between two permanent magnets (neodymium), which were rotated and tilted such that the B$_0$ field was aligned with one of the four possible NV-center orientations. The distance of the magnets was adjusted in order to reach the working magnetic field strength B$_0$ (in this work $\sim$88 mT). Initialization of the NV ensemble was achieved by using a 532 nm laser (Laser Quantum Opus 532) at a power of around 200 mW at the diamond position. The laser light was focused on the diamond by a Plano-Convex Lens (LA1986-A-M, Thorlabs) in a total internal reflection geometry. The laser pulses were controlled by an acousto-optic modulator (Gooch and Housego, model 3260-220) with pulse durations of 5 $\mu$s. For quantum control of the NV centers, microwave pulses were directly synthesized with an arbitrary waveform generator (AWG5202, Tektronix) allowing precise control of phase and frequency, then amplified using a microwave amplifier (ZHL-100W-52 S+, Mini-Circuits). A homebuilt microwave loop was utilized to deliver microwave to the sample.  The diamond was glued to a thin glass cover slide (48393-026, VWR) and a 6 mm glass hemisphere (TECHSPEC\textsuperscript{\textregistered} N-BK7 Half-Ball Lenses, Edmund Optics) was glued to the other side, in order to improve the light collection. This assembly was then glued to a 30 mm cage plate (CP4S, Thorlabs) and mounted above two condenser lenses (ACL25416U-B, Thorlabs) that collected and collimated PL to an avalanche photodiode (A-CUBE-S3000-03, Laser Components GmbH). A long-pass filter (Edge Basic 647 Long Wave Pass, Semrock) placed between the last condenser lens and the photodiode was used to remove the excitation wavelength from the PL light. 
The photo-voltage was digitised with a data acquisition unit (USB-6281 DAQ, National Instruments). A homebuilt wire coil placed above the diamond delivered the reference RF signals which were synthesized by an arbitrary waveform generator (RIGOL LXI DG1022). The electron spin resonance (ESR) frequency measured from the dip in PL (see \zfr{fig1} c)) was used to determine the magnetic field strength and the NV resonance frequency to perform a Rabi experiment, which then determined the $\pi$/2 pulse duration for the spinlock sequences. A thorough optimization of the MW delivery is of crucial importance for the spinlock experiment to be successful and is mainly accomplished by finding the conditions which minimize the damping of the Rabi oscillations. This indicates that the MW field across the NV ensemble is homogeneous enough to achieve an efficient spinlock (\zfr{rabi}).

\begin{figure}[ht]
  \centering
  {\includegraphics[width=0.45\textwidth]{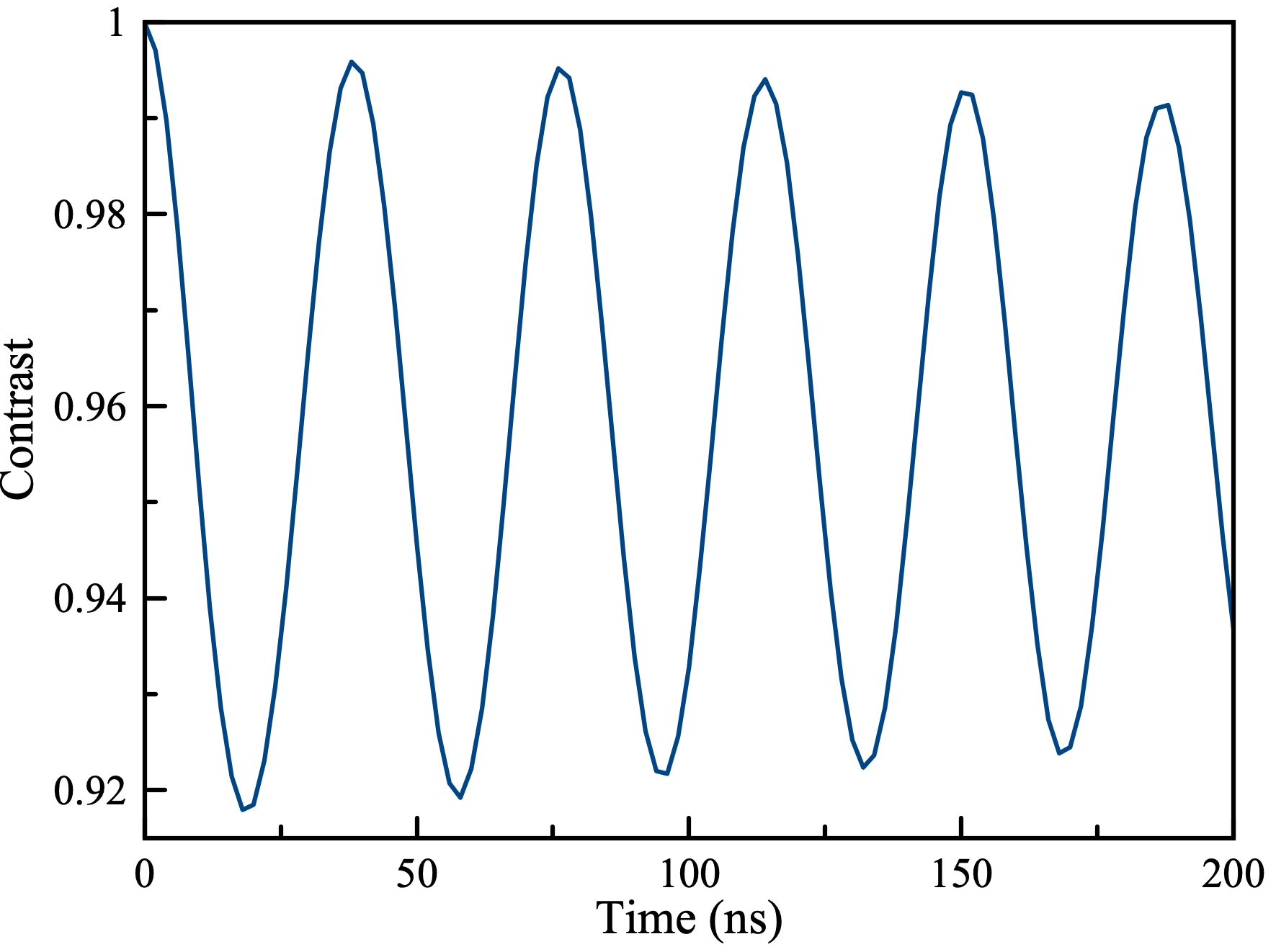}}
  \caption{\textbf{A typical Rabi experiment with NV-ensembles.} The typical contrast of the Rabi oscillations is around 7-8$\%$. This experiment allows to calibrate the pulse length but also to optimize the homogeneity of the MW field throughout the NV ensemble.}
 \zfl{rabi}	
\end{figure}

\textbf{Fittings of the dynamics data. }
Each dataset in \zfr{fig4} a), c) and e) has been normalized to the contrast detected after a strong $\pi$ pulse. The depicted decays start from only a fraction of the available NV polarization because of a fast polarization loss during the first few microseconds spinlock time (not shown). Such a behaviour is likely due to the low MW amplitude of the driving field which causes inefficient spin locking and consequent fast NV T$_{1\rho}$ relaxation. The data points for ``unmatched HH'' in \zfr{fig4} a), c) and e) were fit with a double exponential function whose slower components were chosen to represent the NV relaxation rate constants $k_{T_{1\rho}}$. In particular, for $^{1}$H and $^{19}$F,  $k_{T_{1\rho}}\approx4500 s^{-1}$, whereas for $^{13}$C $k_{T_{1\rho}}\approx9200 s^{-1}$.
In order to maintain constant experimental conditions for comparison, the unmatched HH decay curves had to be acquired at slightly shifted spinlock amplitude (next to the matching point), this causes an overestimation of the polarization transfer rates by 3, 8 and 25$\%$ respectively for $^{13}$C, $^{1}$H and $^{19}$F. 

\section*{Appendix 2: Simulation of the spectral lineshape for a spinlock experiment. }
\begin{figure}[t]
  \centering
  {\includegraphics[width=0.45\textwidth]{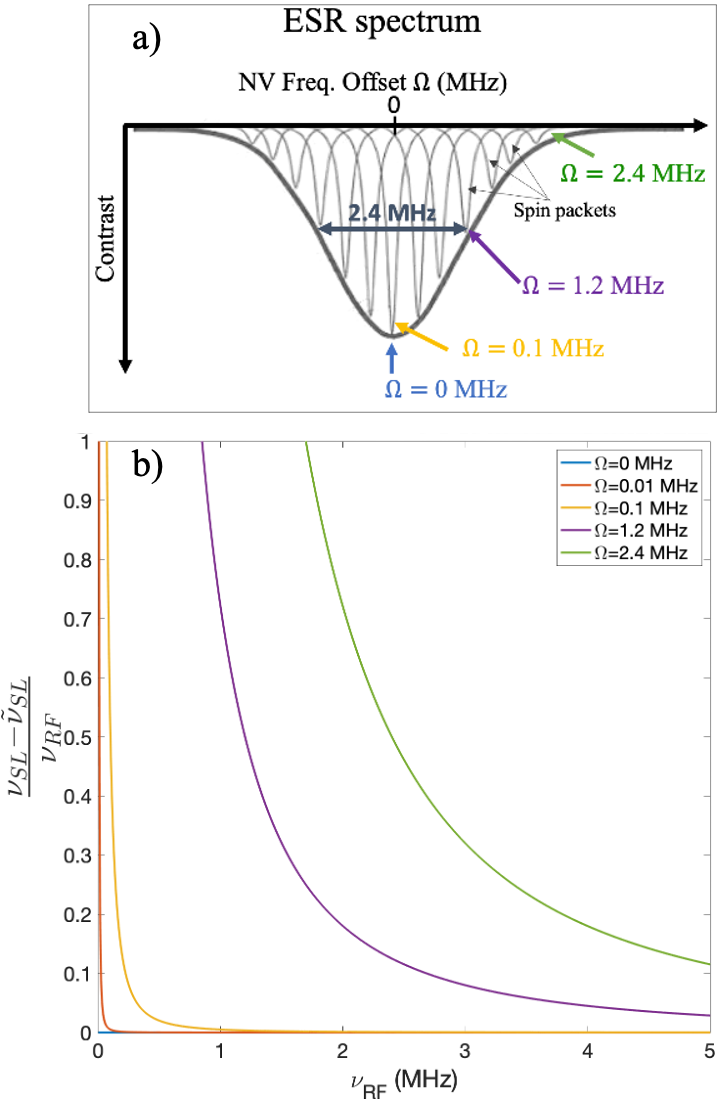}}
  \caption{\textbf{Contribution of the off-resonant NV spin packets to the linewidth of the depolarization dip.} a) Schematic representation of the ESR spectrum with the different NV spin packets resonating at different frequencies  within the ESR line. b) Dependency of the relative spinlock amplitude shift on the RF/Larmor matched frequency. The dependency is plotted for different detunings from $\Omega$=0 to 2.4 MHz.}
 \zfl{linewidth}	
\end{figure}

\textbf{Contribution of the off-resonant NV centers to the linewidth of the NV depolarization dips.}
In this chapter, we derive a formula to predict the contribution of the off-resonant NV spin packets to the linewidth of the depolarization dips, relative to the RF frequency we aim at matching with the spinlock amplitude. 
We recall that during the spinlock experiments reported in \zfr{fig1}, the reference RF is set at a constant frequency $\nu_{RF}$ and the spinlock MW amplitude $\nu_{SL}$ is swept so that matching to the NV Rabi frequencies $\nu_{1NV,i}=\sqrt{{\Omega_i}^2+{\nu_{SL}^2}}$ can occur, where $\Omega_i$ correspond to the offset frequencies of the NV centers within the ensemble. Thus, for perfectly on-resonance NV centers ($\Omega_0=0$), the spinlock amplitude at which the matching condition is satisfied will simply be: $\nu_{SL}=\nu_{RF}$, whereas, for off-resonant NV centers, it will undergo a shift: $\tilde{\nu}_{SL}=\sqrt{\nu_{RF}^2-\Omega_i^2}$. Such a shift will affect the linewidth of the observed depolarization dip, however differently depending on the RF frequency that is matched. By using the following formula (plotted in \zfr{linewidth}), it is  possible to calculate such a dependency:
$\frac{\nu_{SL} -\tilde{\nu}_{SL}}{\nu_{RF}}=\frac{\nu_{RF}-\sqrt{\nu_{RF}^2-\Omega^2}}{\nu_{RF}}=1-\sqrt{\frac{\nu_{RF}^2-\Omega^2}{\nu_{RF}^2}}\approx\frac{1}{2} \left( \frac{\Omega^2}{\nu_{RF^2}} \right)$, since $\sqrt{1-x}\approx1-\frac{x}{2}$, for small $x$ ($0 \leq x \ll 1$), i.e. for $\Omega \ll \nu_{RF}$.

\textbf{Simulation of the ensemble spinlock lineshapes.}
The depolarization peaks reported in \zfr{fig1} e) have been calculated by summing up all single contributions (small peaks underneath) corresponding to slightly shifted matching conditions due to off-resonant NV centers (``spin packets'' having different ESR resonance offset $\Omega_i$). A Matlab script was coded to perform the calculation according to the following scheme: for each $\Omega_i$ within the ESR line (see \zfr{linewidth}), a weighting factor $L$ is calculated corresponding to a normalized Lorentzian intensity having same linewidth $LW$ as the experimental ESR line, see \zfr{fig1} c). Then, $L_i=\frac{1}{1+x^2}$, where $x=\frac{\Omega_i}{LW/2}$ and $LW=2.4$ MHz.
Such a factor is then multiplied to the $ith$ depolarization dip $dip_i$ again utilizing a Lorentzian function as follows: $dip_i=-\frac{1}{1+x_{dip}^2}\times L_i$, where $x_{dip}=\frac{\nu_{SL}-|\tilde{\nu}_{SL}|}{LW_ {dip}/2}$ and $LW_{dip}=60$kHz has been fit to best reproduce the line broadening observed in the experiments.
The lineshape of the depolarization dips by the PROPI experiment reported in \zfr{fig3} e) has also been simulated following the same approach for the matched $^{1}$H case (light blue dips) and $^{19}$F (dark blue dips). In the figure, the intensity of the small dips corresponding to single off-resonant contributions has been multiplied by a factor 3 only for a better visualization.

\section*{Appendix 3: Raw data}
In this section the raw data for the NV depolarization detection by PROPI and NOVEL experiments are reported. As visible in the following figures (Fig. 9-11), such experiments show a typical baseline where the overall contrast increases with the spinlock amplitude. This is due to the fact that, after the strong excitation of the NV spins by the first $\pi/2$ pulse, the spinlock amplitude is stepwise increased and the stronger the spinlock pulse becomes the more efficiently it is able to lock the phase of the NV spins on the transverse plane. In \zfr{fig1}, the baseline removal has been accomplished by subtracting a dataset of the same experiment performed with no RF signal. In \zfr{fig3} c) and d) the baseline has been removed by subtracting a polynomial function fitting to the data of the spectral ranges that contain no peaks. In the following experiments shown, the spinlock amplitude is reported in units of Volt peak-peak(Vpp) as directly read out from our AWG.

\section*{Appendix 4: Simulation of polarization buildups}
All polarization curves reported in \zfr{fig6} have been simulated implementing in a Matlab code, the model schematically depicted in \zfr{fig5}. In particular, the code solves the following system of ordinary differential equations:
\begin{equation*}
    \begin{cases}
    \frac{dP_S}{dt} = -k_{SI}P_S+k_{IS}P_I-k_{T_{T_{1n}\rho}}P_S;\\
    \frac{dP_I}{dt} = -k_{IS}P_I+k_{SI}P_S-k_{T_{1n}}P_I-k_{T_{SD}}P_I 
    \end{cases}.
\end{equation*}
for a time duration dictated by the optimal spinlock time that was experimentally determined (e.g. 65 $\mu$s for $^{19}$F or 5 $\mu$s for $^{13}$C). Then, depending on the number of pulse sequence repetitions that we wanted to simulate, the code was looped resetting only the electron spin  polarization $P_S$ = 1, at each time.
Rate constants, where possible, have been measured with the experiments reported above. $T_{1n}$ has been considered to be on the order of 100 s for $^{13}$C inside the diamond and 1 s for both $^{1}$H and $^{19}$F\cite{ajoy_hyperpolarized_2019,tetienne_prospects_2021}.
The spin diffusion time constant $T_{SD}$, has been derived using a formula adopted from simple random walk theory $T_{SD}\approx r^{2}/(4\cdot D)$, where $r$ is the average displacement and $D$ the spin diffusion constant. This gives an estimation on how long it takes for the polarization generated in one point of the sensing area to leave the detection volume of the NV-center. $D$ = 700 nm$^2 s^{-1}$ and $r$= 10 nm for both $^{19}$F and $^{1}$H and  $D$ = 0.0035 nm$^2 s^{-1}$ (\cite{broadway_quantum_2018}) and $r$ = 1 nm for $^{13}$C have been considered.
\added[id=RR]{The inverted depolarization dip observed for $^{13}$C during the NOVEL experiment and reported in \zfr{fig3} can be rationalized by means of our simplified rate equations model too. 
In the same manner as it was done for \zfr{fig6}, in \zfr{fig12} the $\%$ of NV spin polarization has been reported vs the number of pulse sequence repetitions. At start of each pulse sequence, the NV spin is reset to 100 $\%$ polarization, whereas the nuclear spin polarization is allowed to build up during a 40$\mu$s spinlock time (the same pulse length used in the experiment of \zfr{fig3}).  We note that, due to the decay of the NV polarization caused by the $T_{1\rho}$ relaxation, at a certain point of the dynamics the nuclear spin bath reaches a polarization higher than the one of the NV spins. Thus, after reaching this point and for the subsequent repetitions, the spinlock pulse can partially work in the opposite direction letting some polarization to flow back from the nuclear to the NV spins.  This will occur during the last fraction of the spinlock duration, that is after the NV is relaxed to a polarization level lower than the nuclear one. When we compare the signal for the matched HH condition with the one where no condition is met, (as we do experimentally when we compare the intensity of the dip to the baseline), we see that the mechanism described above results in an inversion of the “depolarization dip” and this effect can become more severe the shorter the $T_{1\rho}$ time constant and/or the longer the spinlock time.} 
 \begin{figure}[htbp]
 \centering
  \setlength{\unitlength}{\textwidth} 
  {\includegraphics[width=0.5\textwidth]{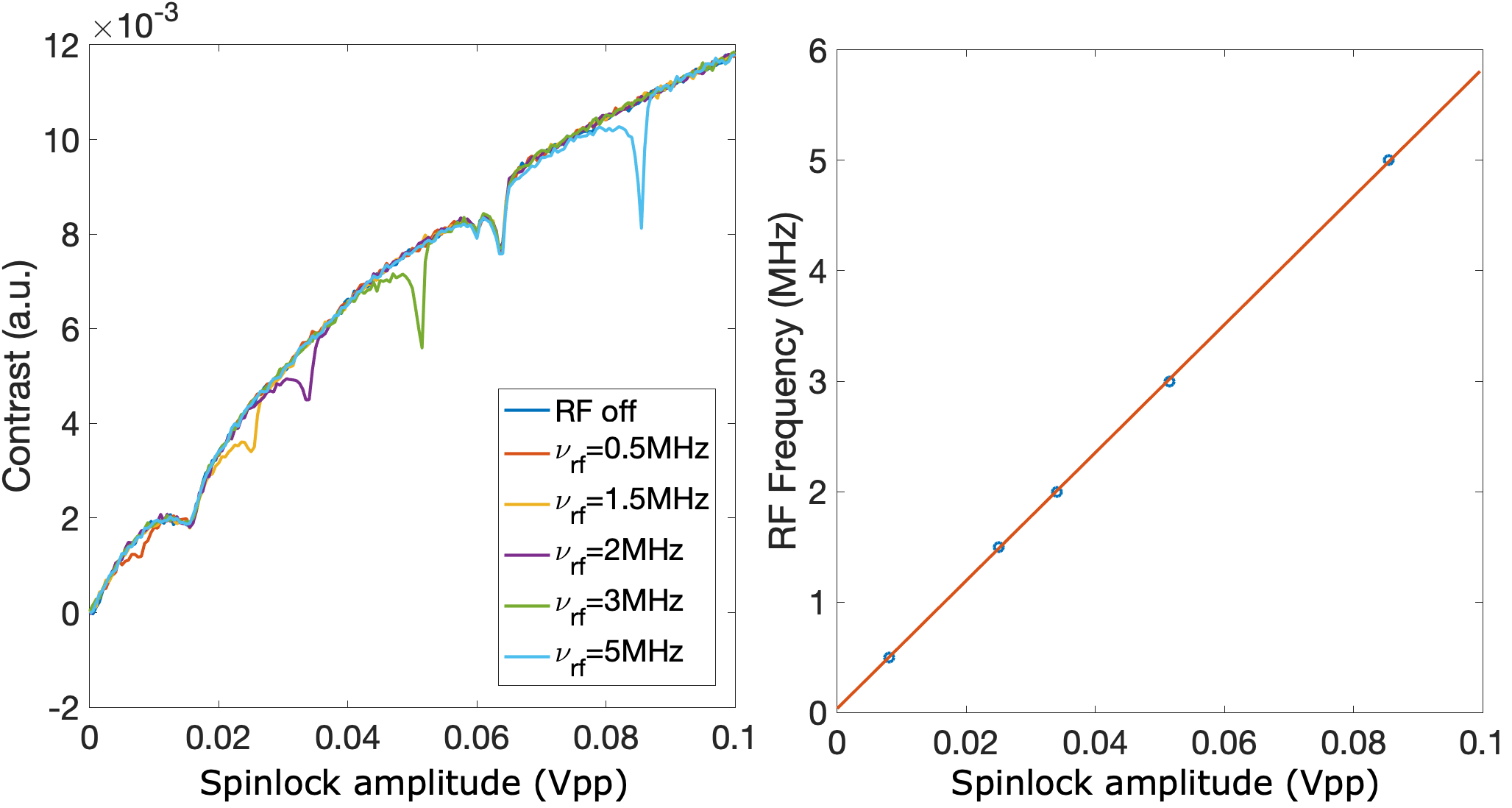}}
  \caption{\textbf{Calibration of the spinlock experiment by means of an external RF signal.} Left: the signal has been acquired by averaging each point 5000 times and the whole sweep repeated 3 times. Right: The spinlock amplitudes at which the depolarization dips appeared upon application of the different $\nu_{RF}$ have been read out and plotted. A linear fit is showed in red.}
 \zfl{figb1}	
\end{figure}
\begin{figure}[htbp]
  {\includegraphics[width=0.5\textwidth]{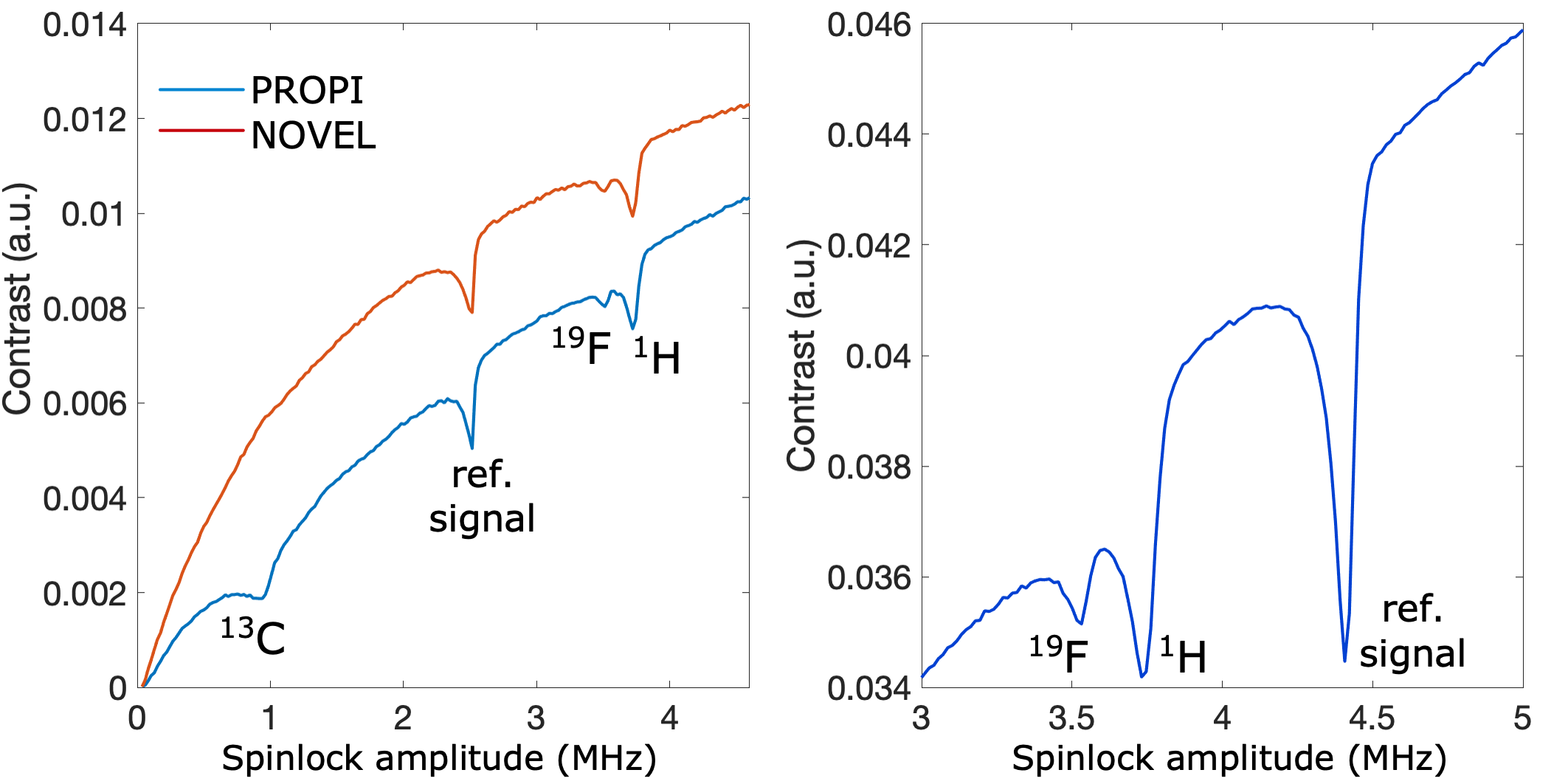}}
  \caption{\textbf{Raw data for the SAM layer on the diamond.} Left: 5000 averages for 20 sweeps. Right: 5000 averages for 250 sweeps.}
 \zfl{figb2}	
\end{figure}
\begin{figure}[htbp]
  {\includegraphics[width=0.5\textwidth]{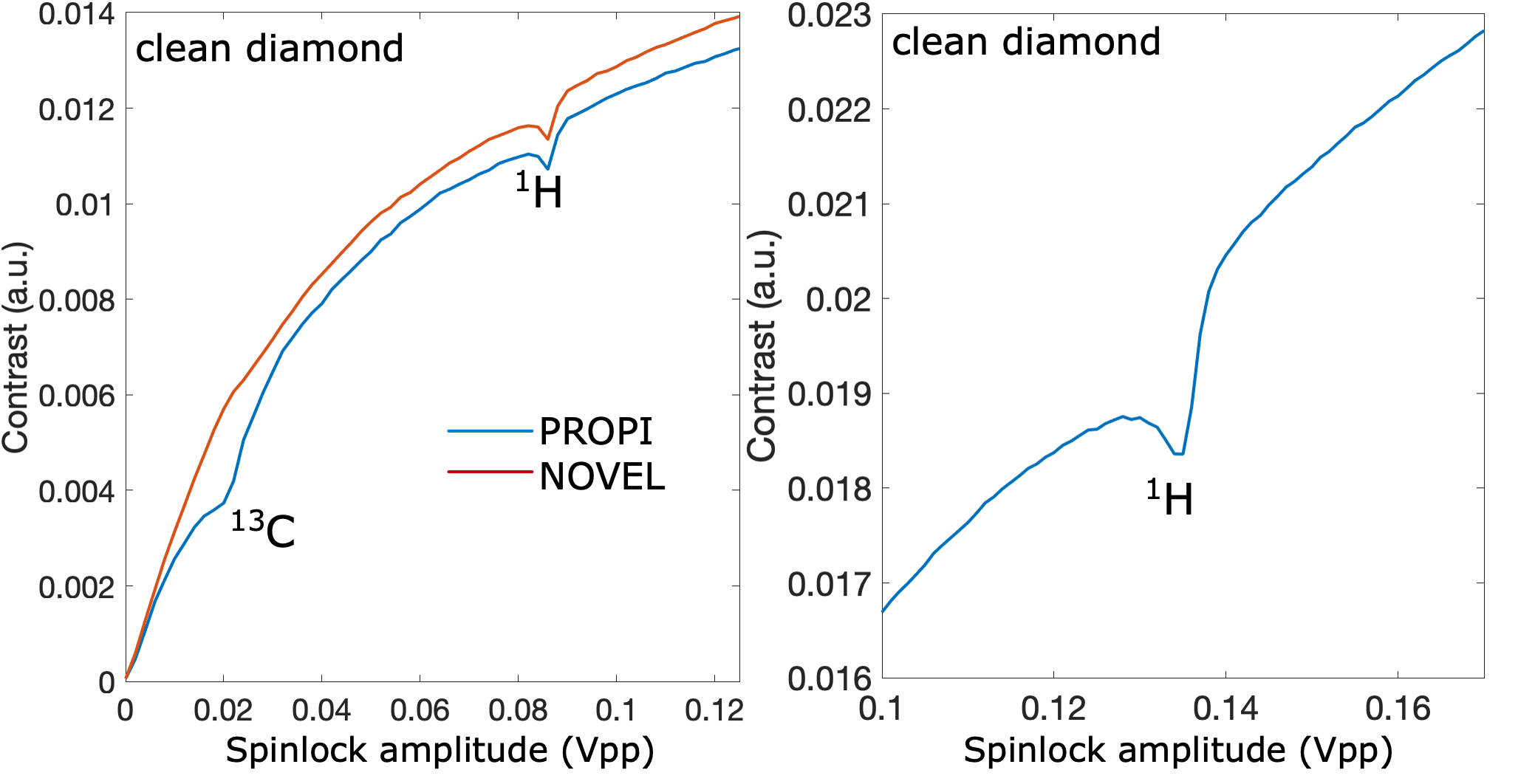}}
  \caption{\textbf{Raw data for the clean diamond.} Signal acquisition: 5000 averages for 20 sweeps. }
 \zfl{figb3}	
\end{figure}
\begin{figure}
  {\includegraphics[width=0.45\textwidth]{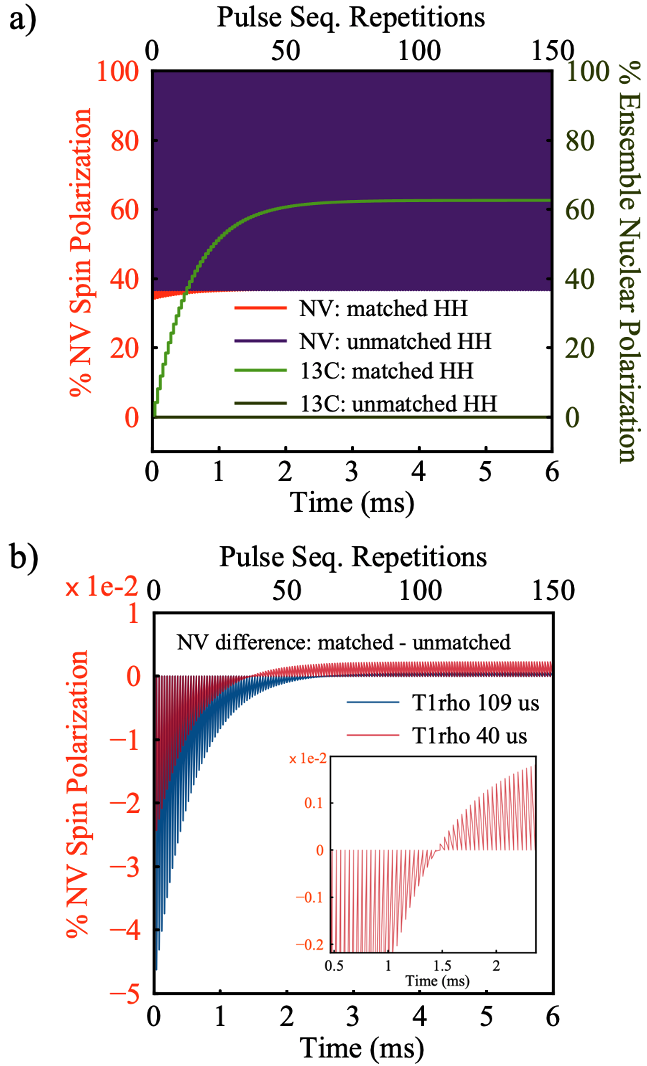}}
  \caption{\added[id=RR]{\textbf{Inversion of the depolarization dip at $^{13}$C matching observed during NOVEL.} a) Simulation of the NV (red) and nuclear (green) spin polarization as a function of the pulse sequence repetitions for the cases: i) when the polarization transfer rates are activated (matched HH) ii) when the polarization transfer rates are off (unmatched). Spinlock pulse duration is 40 $\mu$s, all other rates as in \zfr{fig6}. b) Difference between calculated NV polarization for the cases when the HH condition is matched and non-matched. The decay at the first pulse sequence corresponds to the full intensity of the depolarization dip (the same observed during a PROPI experiment). After some repetitions of the pulse sequence the dip gets less and less intense until it inverts.The simulation is shown for two different  $T_{1\rho}$ time constants (red 40 $\mu$s and blue 109 $\mu$s).}}
 \zfl{fig12}
 \end{figure}
\clearpage
\bibliography{bibliography.bib}
\end{document}